\documentclass[10pt,twocolumn,letterpaper]{article}

\usepackage{icb}
\usepackage{times}
\usepackage{epsfig}
\usepackage{graphicx}
\usepackage{amssymb}
\usepackage{multirow}
\usepackage{multicol}
\usepackage{mathtools}          %
\DeclareRobustCommand{\frac}[3][0pt]{%
  {\begingroup\hspace{#1}#2\hspace{#1}\endgroup\over\hspace{#1}#3\hspace{#1}}}
\usepackage[norule,symbol,perpage]{footmisc}

\usepackage{fancyhdr}

\icbfinalcopy %

\ificbfinal\fi

\allowdisplaybreaks

\fancypagestyle{plain}{
	\fancyhf{} % clear all header and footer fields
	\fancyhead[C]{\textcolor{red}{Richard Plesh et al., ``Fingerprint Presentation Attack Detection utilizing Time-Series, Color Fingerprint Captures'',\\
	Proc. of 12th IAPR International Conference on Biometrics (ICB), (Crete, Greece), June 2019.}}
	\fancyfoot[R]{}

}

\begin{document}

\title{Fingerprint Presentation Attack Detection utilizing Time-Series, Color Fingerprint Captures}

\author{Richard Plesh\\
Clarkson University\\
Potsdam, NY, USA\\
{\tt\small pleshro@clarkson.edu}
\and
Keivan Bahmani\\
Clarkson University\\
Potsdam, NY, USA\\
{\tt\small bahmank@clarkson.edu}
\and
Ganghee Jang\\
Clarkson University\\
Potsdam, NY, USA\\
{\tt\small gjang@clarkson.edu }
\and
David Yambay\\
Clarkson University\\
Potsdam, NY, USA\\
{\tt\small yambayda@clarkson.edu}
\and
Ken Brownlee\\
Silk ID Systems Inc.\\
Santa Clara, CA, USA\\
{\tt\small ken@silkid.com}
\and
Timothy Swyka\\
Precise Biometrics\\
Potsdam, NY, USA\\
{\tt\small tim.swyka@precisebiometrics.com}
\and
Peter Johnson\\
Precise Biometrics\\
Potsdam, NY, USA\\
{\tt\small peter.johnson@precisebiometrics.com}
\and
Arun Ross\\
Michigan State University\\
East Lansing, MI, USA\\
{\tt\small rossarun@cse.msu.edu}
\and
Stephanie Schuckers\\
Clarkson University\\
Potsdam, NY, USA\\
{\tt\small sschucke@clarkson.edu}
}

\maketitle

\begin{abstract}

Fingerprint capture systems can be fooled by widely accessible methods to spoof the system using fake fingers, known as presentation attacks. As biometric recognition systems become more extensively relied upon at international borders and in consumer electronics, presentation attacks are becoming an increasingly serious issue. A robust solution is needed that can handle the increased variability and complexity of spoofing techniques. This paper demonstrates the viability of utilizing a sensor with time-series and color-sensing capabilities to improve the robustness of a traditional fingerprint sensor and introduces a comprehensive fingerprint dataset with over 36,000 image sequences and a state-of-the-art set of spoofing techniques. The specific sensor used in this research captures a traditional gray-scale static capture and a time-series color capture simultaneously. Two different methods for Presentation Attack Detection (PAD) are used to assess the benefit of a color dynamic capture.  The first algorithm utilizes Static-Temporal Feature Engineering on the fingerprint capture to generate a classification decision. The second generates its classification decision using features extracted by way of the Inception V3 CNN trained on ImageNet. Classification performance is evaluated using features extracted exclusively from the static capture, exclusively from the dynamic capture, and on a fusion of the two feature sets. With both PAD approaches we find that the fusion of the dynamic and static feature-set is shown to improve performance to a level not individually achievable.
\end{abstract}

\section{Introduction}

Biometric characteristics are valuable for recognizing individuals since they are distinctive between individuals and are inherent to every person. Since biometric characteristics can be publicly observed, systems cannot assume security from secretiveness. For example, latent fingerprints can be easily lifted from many different surfaces which can be used to create fake biometrics and attack the system, also called presentation attacks (PA). Therefore, unlike traditional password systems, the presentation of a matching biometric characteristic cannot be the only requirement for recognition. A second step is needed to verify that the biometric characteristics are being presented by an authorized, live, and present source, and this is called Presentation Attack Detection (PAD) \cite{schuckers_presentations_2016}.

The current landscape of PAD methods consists of  hardware-based and software-based approaches. The former requires the collection of further finger information via additional sensor hardware, such as spectral absorption, skin temperature, and pulse oximetry \cite{adair_nixon_spoof_2007}. Software-based approaches do not require additional hardware because they utilize the information contained in the digital image captured by the sensor \cite{schuckers_presentations_2016}. Software approaches are further divided into static and dynamic properties. Static properties come from a single fingerprint image, while dynamic are derived from a time-series capture. Researchers have developed many different methods of extracting features from either static or dynamic captures.

Researchers from the University of Utah and MIT studied the effect of pressure on the hemodynamic state of the fingertip. They found that the application of force produced blood perfusion coloration patterns beneath the fingernail that were common among all people in general \cite{mascaro_common_2006}. Previous work by Wei-Yun Yau, Hoang-Thanh Tran, Eam-Khwang Teoh, and Jian-Gang Wang demonstrated the viability of detecting fake fingers using finger color change by developing and testing a method of measuring blood perfusion when a finger is applied to a sensor \cite{yau_fake_2007}. The data set used for testing was very small, was not generated using a commercial fingerprint sensor, and contained only one type of spoof finger. The research also does not evaluate any other dynamic changes that could be used for PAD. Our research addresses each of these shortcomings by developing and applying a collection of dynamic color features to a large and diverse data set generated using a sensor from SilkID Systems Inc.

Live fingerprints have unique properties in their fingerprint ridge signals due to the buildup of perspiration moisture. Previous publications have presented a low-cost software approach for time-series PAD that can utilize the perspiration phenomena in captures \cite{parthasaradhi_time-series_2005}, \cite{decann_novel_2009}.  Since perspiration is a relatively difficult psychological phenomenon to spoof, multiple methods have been developed for identifying perspiration for the purpose of spoof detection. In \cite{tan_liveness_2006}, perspiration is detected using a wavelet transform on a single image (static capture) fingerprint ridge signal. In \cite{parthasaradhi_time-series_2005}, perspiration is detected using various measures of change in the ridge signal over a multi-frame fingerprint capture (dynamic capture). 

The research could benefit from a more diverse data set however.
The proposed algorithms also have practical limitations because time-series data rarely remain uniform enough to generate a meaningful difference image. This irregularity is due to the finger deforming and shifting during the capture period.  In order to apply these techniques more successfully, we generate the feature value difference between frames, rather than extract features from a difference image, and test using a large data set.

An intensity-based approach has been applied by performing an image histogram of the fingerprint capture center. The ``square area" of analysis is manually defined to avoid large scars and encompass an interesting part of the fingerprint \cite{tan_liveness_2005}.
 This method has potential robustness advantages since the feature values should be resistant to skewing caused by smudged or shifting prints. In our research we utilize an automated and dynamic definition of the area of analysis, device-specific algorithm tuning, and a more diverse data-set for training. Additionally, time series features in this paper take into account shifting of the finger over the capture period by normalizing for each frame by ridge signal as well as foreground and full fingerprint ridge.

Local textural features have also been used for liveness detection \cite{ghiani_textural_2015}. These descriptors perform analysis on small patches of the fingerprint surface, in contrast to the global descriptors that use the entire image.

The analysis of skin distortion for spoof detection was performed in \cite{antonelli_fake_2006}, \cite{zhang_fake_2007},  \cite{cappelli_modelling_2001}. The technique in \cite{antonelli_fake_2006} requires a user to apply finger in a twisting motion with pressure while the sensor conducts a time-series capture. The research found that fake fingers are more rigid than human skin, with much less distortion.  An approach that could measure distortion from a normal finger application (without twisting motion) would have practical ergonomic benefits for the user.

While previous work has explored dynamic features, this paper extends prior work by including advanced dynamic features based on color, evaluation on a large data set of over 36,000 image sequences of live fingerprints and a diverse set of spoof types, and feature-level fusion between dynamic and static features generated simultaneously.

This paper explores the utility of a color time-series fingerprint capture to fingerprint Presentation Attack Detection (PAD). We present the performance of state-of-the-art PAD algorithms based on a traditional gray-scale static (one frame) fingerprint capture, the state-of-the-art PAD algorithm for a color time-series (dynamic) capture, and a fusion of the two approaches. The dynamic PAD algorithms consist of novel features fused together with those from previous research while the static PAD algorithms are based on techniques introduced in prior papers. For training and testing of the algorithms we compose a diverse data set of live and spoof fingerprints using state-of-the-art spoofing techniques.

\section{Dynamic and Color Fingerprint}
While a static fingerprint capture gives us valuable liveness information, certain attributes of vitality cannot be captured effectively with a single gray-scale fingerprint capture. A dynamic color capture from a fingerprint sensor has the potential to provide access to difficult to spoof physiological signs of vitality, such as the displacement of blood in the finger when pressed. A time-series PAD method could also provide additional information on the spread of perspiration and the deformation of the finger over time. By utilizing the capability of the sensor to simultaneously capture a static and color time-series fingerprint (called hereon, fast frame rate or FFR), we explore the value of dynamic information to PAD performance. 

State-of-the-art dynamic fingerprint capture PAD and baseline static fingerprint PAD are tested in isolation to evaluate the relevancy of the information in each capture to liveness detection. Last of all, the features extracted from each type of capture are fused together and evaluated to determine the novelty of the dynamic information to PAD. We hypothesize that a relatively high performance from a fusion method is indicative that each method is capturing a significant amount of unique information.

The sensor used was manufactured by Silk ID Systems Inc. and had specialized firmware loaded to perform the FFR functions.  This sensor is a conventional optical fingerprint scanner utilizing dark-field, frustrated-total-internal-reflection as its imaging technique.  The touch surface is a glass prism.  A conventional CMOS Image Sensor of 2 megapixels resolution and standard color filter is used to capture the image.  Four white LEDs illuminate the finger.  When an IR finger detection circuit detects the presence of an object on the touch surface, a burst of 8 images are captured at approximately 8 -10 frames per second.

\begin{figure}[htbp]
\begin{center}
  \includegraphics[width=0.7\linewidth]{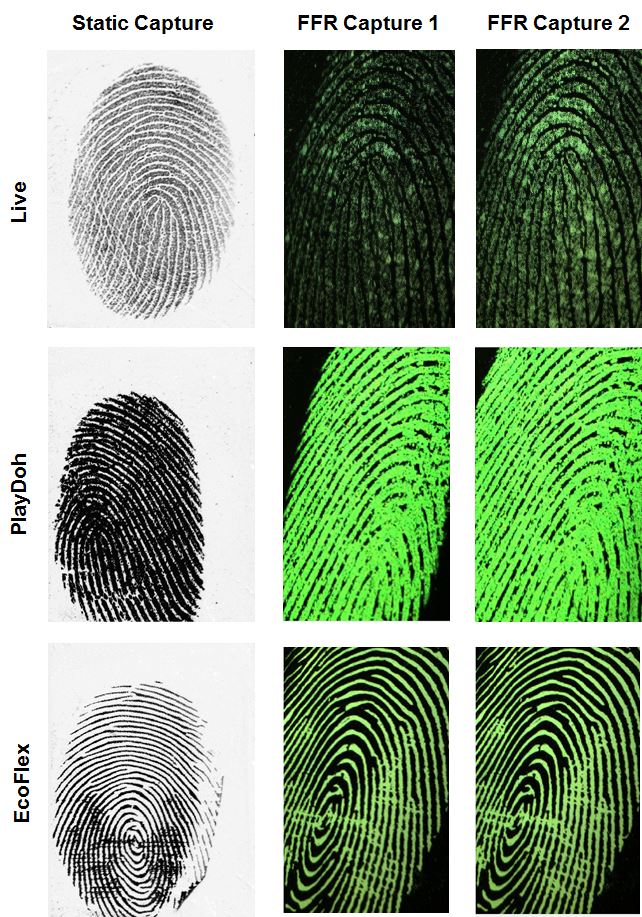}
    \caption{Static Image and selected frames from 3 different SilkID Captures from Live subject, PlayDoh PA, and EcoFlex PA}
  \label{fig:silk_capture_example}
\end{center}
\end{figure}

The images analyzed for the PAD are in their raw configuration to preserve the highest resolution.  Some inherent distortion due to the optical configuration is present.  This is later corrected by software to yield a geometrically-correct image for fingerprint matching.   There are twice as many green pixels in the standard Bayer pattern of CMOS image sensors, yielding the strong green appearance in the raw images. Figure \ref{fig:silk_capture_example} presents three different captures from the SilkID sensor. Only two frames of the FFR capture are shown. Also, in this paper we refer to the foreground as the portion of the scanner image that contains the fingerprint and the background as the surrounding blank space.

\section{Feature Extraction}
This section describes the details of the static and dynamic feature extraction process. We employed several hand engineered feature extraction methods as well as Convolutional Neural Networks (CNNs) for feature extraction. 

\subsection{Optimum Frame Selection}

The sensor captures a sequence of 8 frames to acquire the finger of the subject. Since the PAD approach presented in this paper calculates dynamic liveness character using the differences between only two frames in the time-series, a method is needed to select the best two frames of the captured sequence. One frame near the beginning, and one the end. We found that the first few frames could have inconsistent amounts of finger surface in the capture while the ending few frames had very little dynamic character since the finger was largely settled. 
In other words, selecting an earlier beginning frame would beget dynamic change at the expense of consistency. To solve this trade-off, the first frame is chosen as the earliest non-blank frame, while the second is at least 625msec from the first.

For the remainder of this paper, we use ``1'' to denote the beginning frame chosen by the algorithm, and ``2'' to denote the ending frame chosen by the algorithm.  The time between frames is less than one second.

\subsection{Dynamic Features}

\subsubsection{Dynamic Color features}
\begin{figure}[htbp]
\begin{center}
  \includegraphics[width=.8\linewidth]{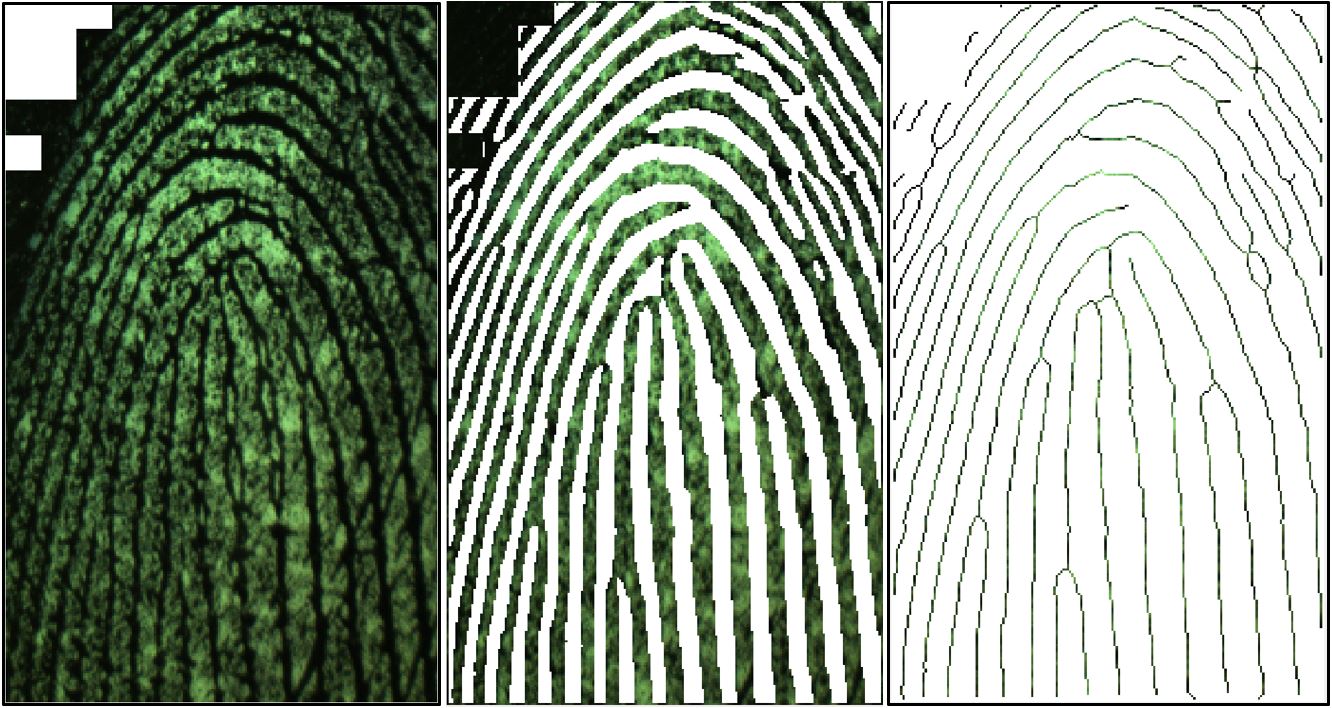}
    \caption{Fingerprint foreground, fingerprint ridge pixels, and the pixels used for generating the fingerprint ridge signal (left to right)}
  \label{fig:three_pixel_groupings}
\end{center}
\end{figure}

When a live finger is pressed onto a sensor, a unique color change occurs, possibly due to the displacement of blood. The same mechanism makes our skin appear white when pressed up against a window. This dynamic character can be quantified and used for spoof detection. We have several different methods for extracting dynamic color information from the Fast Frame Rate (FFR) capture. In addition, three different portions of the fingerprint image are targeted for each extraction method and color ratio. The three portions are the fingerprint foreground, the fingerprint ridges, and the fingerprint ridge signal as shown in figure \ref{fig:three_pixel_groupings}. The foreground pixels are calculated by way of a mask that blocks out areas of that do not contain the fingerprint. The identification of the ridge signal pixels and generation of the ridge signals is performed via the fingerprint enhancement routine described in previous work \cite{tan_liveness_2006}.

As the finger is applied to the sensor, we expect the values for each color channel to increase over time. In order to specifically target the change in finger color to white from the displacement of blood, we measure the change in green/red and blue/red ratio images between image 1 and image 2. The green/red and blue/red ratio images for the foreground pixels and ridge-signal are created via element-wise division between the images from each channel as shown in equation \ref{eqn:FN_CRI}. Since the ridge signals may be shifted slightly over the course of the capture, a realignment routine uses correlation analysis to realign the signals. Also, the calculations on the ridge pixels are performed slightly different due to the difficulty in aligning the fingerprint from frame 1 to that of frame 2. Rather then using an element-wise division, the mean intensity is calculated on the raw color channels, and then divided to generate a color ratio measure as shown in equation \ref{eqn:FN_CRM}. In both cases a 0.001 term is added to the denominator to prevent dividing by zero. The change in the ``ratio-image" is calculated using difference in mean, ratio of means, and the square root of the sum of squared differences as shown in equation \ref{eqn:diff}, \ref{eqn:ratio}, and \ref{eqn:sumsquare} respectively.  We expect the live subject to have a green/red or blue/red color ratio change greater in live over spoof. 
To explore other potential liveness phenomena experimentally, we perform dynamic change calculations on the green/blue ratio and each of the color channels individually as well. The outcomes of equations \ref{eqn:diff}, \ref{eqn:ratio}, \ref{eqn:sumsquare}, and \ref{eqn:sequence_euclid} are used as features for the classifier. Since there are multiple color channels and finger areas targeted, there are 12 features extracted from the fingerprint foreground, 12 features extracted from the ridge signal, and 7 features extracted from their ridge pixels.

   \begin{eqnarray}
    \label{eqn:FN_CRI}
    FN_{CRI} &=& FN_{clri}\oslash(FN_{clrj} +0.001) \\
    \label{eqn:FN_CRM}
    FN_{CRM} &=& \frac[10pt]{\overline{FN_{clri}}}{\overline{FN_{clrj}}} \\ 
    \label{eqn:diff}
    diff &=& \overline{F2_{ms}} - \overline{F1_{ms}}  \\ 
    \label{eqn:ratio}
    ratio &=& \frac[10pt]{\overline{F2_{ms}}}{\overline{F1_{ms}}}\\
    \label{eqn:sumsquare}
    sumsquare &=& \sqrt{\sum[F2_{ms}-F1_{ms}]^{2_{(ew)}}} \\
    \label{eqn:sequence_euclid}
    Sequence_{euclid} &=& \sqrt{ R_{diff}^2 + G_{diff}^2 + B_{diff}^2 }
\end{eqnarray} 
where,\\\\
$FN_{clr}$: Image pixels from the Nth frame position (1: First Frame, 2: Last Frame) of channel clr (red, green, blue). \\
$FN_{CRI}$: Color Ratio Image for the Nth frame. \\
$FN_{CRM}$: Color Ratio Measure for the Nth frame. Needed for the ridge pixel features. \\
$FN_{ms}$: Measure from the Nth frame. \\
${(ew)}$: Signifies element-wise operation \\
$C_{diff}$: Difference in mean channel intensity between frames for channel color. Example: $R_{diff}$, $G_{diff}$, $B_{diff}$.\\
$diff$: Feature generated using the difference between two measures from each frame. \\
$ratio$: Feature generated using the ratio of measures from each frame. \\
$sumsquare$: Feature generated using the square root of the sum of the squared differences between the two frames. \\
$Sequence_{euclid}$: Feature generated from the Euclidean distance between the mean intensity values for the three channels.\\\\

\subsubsection{Finger Shifting Features}

In the data set we observed that spoofs generally have more shifting associated with them. We attribute this to the awkwardness of applying a fake finger. We quantify this change and use it as a feature for the classifier.

To detect the motion, we utilize the criteria outlined in equation \ref{eq:delta}. Let’s call it delta image. The squared sum of pixel-wise difference captures the difference between images. The time difference between consecutive frames is 125 msec with marginal drift, so the difference between a neighboring pair can indicate the degree of movement of a finger. By limiting the squared difference to 255, the sum is more indicative of movement rather then color change. 

\begin{eqnarray*}
press(x,y)_{frame(i)} &=& (red(x,y)_{frame(i)} \\
            && {} - red(x,y)_{frame(i+1)})^2
\end{eqnarray*}
\begin{equation}
{Delta\: Image} = \sqrt{\displaystyle\sum_{frame(i)}^{} press(x,y)_{frame(i)}}
\label{eq:delta}
\end{equation}

\subsubsection{Foreground/Background Mask Features}

During our PAD processing we develop a mask to separate the foreground and background of the finger capture. We noticed that the mask itself contains information on liveness. It appears that spoof fingerprints generally have a smaller foreground and larger background as well as more movement in the foreground mask relative to spoofs.
The foreground and background areas are generated via sum of the mask elements as shown in equation \ref{eqn:AN_fore} and \ref{eqn:AN_back} respectively. The ratio of foreground to background in each frame, the ratio change of foreground/background, the difference to foreground/background ratio, and the shift in foreground mask are calculated from equations \ref{eqn:RN_foreback} to \ref{eqn:ms}.

\begin{eqnarray}
	\label{eqn:AN_fore}
    AN_{Fore} &=& \sum (Mask)\\
    \label{eqn:AN_back}
    AN_{Back} &=& \sum (\sim\ Mask) \\
    \label{eqn:RN_foreback}
    RN_{ForeBack} &=& \frac{AN_{Fore}}{AN_{Back}} \\
    \label{eqn:RS}
    RS &=& \frac{R1_{ForeBack}}{R2_{ForeBack}+0.0001}\\
    \label{eqn:delta_back}
    \Delta_{Back} &=& A2_{Back} - A1_{Back} \\
    \label{eqn:ms}
    MS &=& \sum A1_{Fore}\oplus A2_{Fore}
\end{eqnarray}

where,\\
\[
  Mask(n,m) =
  \begin{cases}
                                   1 & \text{if $frame(n,m)$ is fingerprint} \\
                                   0 & \text{if $frame(n,m)$ is empty background} 
  \end{cases}
\]
$AN_{Fore}$: The Area of the Foreground mask. \\
$AN_{Back}$: The Area of the Background mask. \\
$RN_{ForeBack}$: The Ratio of Foreground to Background for Frame N \\
$RS$: Ratio of the Foreground/Background measure for Frame 1 to that of Frame 2 \\
$\Delta_{Back}$: The difference between the background area of Frame 1 and 2 \\
$MS$: A measure of shifting between Frame 1 and 2 using XOR to detect differences in Mask Profile \\\\

\subsubsection{Dynamic Fingerprint Foreground Intensity}
Dynamic foreground features are designed to capture the global intensity changes in the dynamic capture of the fingerprint and are a modification of the features presented in Tan, et, al \cite{tan_liveness_2005}. Since these features do not concern the finger color information specifically, the RGB dynamic capture is converted to grayscale for analysis. The previous paper manually selected a section in the center of the fingerprint capture for analysis. Since a practical implementation of this algorithm would require an automated selection of capture area,
we dynamically determine the foreground of the capture for frequency analysis and normalize accordingly. Foreground selection allows us to observe the intensity information for the entire finger area, rather then just a small section, and improves robustness by diminishing capture variability. An image intensity histogram is then generated from the two selected frames. Static measures are extracted from a single image histogram while dynamic features are generated from the difference between the two histograms. The measures extracted from the histograms needed to be tuned to the specific SilkID sensor characteristics, such as the inversion of black and white. Through this method, 6 features are extracted from the classifier.

\subsubsection{Dynamic Ridge Signal Perspiration Detection}
When a live finger is in contact with a sensor, a dynamic perspiration pattern can be detected in the ridge signal. To extract the temporal changes in the ridge signal due to perspiration we rely on the method described in Parthasaradhi et al\cite{parthasaradhi_time-series_2005}. Just as in this paper, we extract 1 static and 6 dynamic features from the fingerprint ridge signal. These features use the Fourier transform of the ridge signal, total swing in ridge signal, growth ratio of signal peaks, growth of signal mean, percent change in signal variance, dry saturation change, and wet saturation change.

\subsection{Static Features}

\subsubsection{Engineered Feature Extraction}
Our baseline engineered feature algorithm extracts features from only the static image capture. These features include local binary pattern, fingerprint foreground intensity, and multi-resolution fingerprint ridge and valley texture analysis features from previous research. Local binary pattern (LBP) analysis has been commonly used for biometric recognition and spoof detection \cite{ghiani_textural_2015}. 
In this research LBP was performed for two radii (1 and 2 pixels). Fingerprint foreground intensity utilizes the relative frequencies of different intensity levels within the fingerprint capture. Multi-resolution fingerprint ridge and valley texture analysis uses the wavelet transform to decompose the ridge and valley signal into multi-scales. Features are then extracted from each scale to quantify the perspiration pattern. 72 features are generated from LBP analysis, 64 features from fingerprint foreground intensity, 14 features from multiresolution ridge analysis, and 14 features from multiresolution valley analysis.

\subsubsection{Feature Extraction Using Convolutional Neural Networks}

In this work, we employed Inception-V3 CNN model \cite{szegedy2016rethinking} trained on Imagenet dataset \cite{deng2009imagenet} to extract features from the static images. The model is implemented using Keras platform \cite{chollet2015keras}, where each static images is re-sized to 299 by 299 pixels and 2048 bottleneck features are extracted using the CNN model.

\begin{table}[!t]
\renewcommand{\arraystretch}{1.3}
\caption{Summary of collected samples from presentation attack using finger spoofs.}
\label{table:fingerPAdetail}
\centering
\begin{tabular}{|c|c||c|}
\hline

\multicolumn{2}{|c||}{Type of PA attack}  & \multirow{2}{*}{Captures} \\
\cline{1-2} Mold & Material & \\
\hline
\hline
3D & Dragonskin & 258\\
3D & Ecoflex & 267\\
3D & Gelatin & 201 \\
3D & ModelMagic & 258 \\
3D & PlayDoh & 203 \\
3D & SillyPutty & 265 \\
3D & Wood glue & 217\\
\hline
Dental & Ecoflex & 3924 \\
Dental & Gelatin & 1543 \\
Dental & ModelMagic & 1484 \\
Dental & PlayDoh &  3190 \\
Dental & SillyPutty & 1840 \\
Dental & Wood glue & 2378 \\
Dental & Latex BodyPaint & 1245 \\
\hline
N/A & Paper (2D print) & 2893 \\
N/A & Transparent film & 1534 \\
\hline
N/A & Live Finger & 14892 \\
\hline
\hline
\multicolumn{2}{|c||}{Sum All} & 36592\\
\hline
\end{tabular}
\end{table}

\subsection{Data Set}

One of the distinct characteristics of our data set is the variety of fake fingers, both in terms of mold type and the materials which form the fake fingers. In terms of molds, there are 3 distinct categories: 3D molds from 3D printer, dental molds, and 2D printing. 

The 3D molds are created through the use of a 2D fingerprint image. The 2D fingerprint image is binarized and converted into a surface using the binarized grayscale values. For the molds used in this dataset, the ridge values are extruded down and the valley values are kept at a depth of 0. This new surface is re-meshed using meshmixer to ensure the integrity of the mesh and re-sized based on the original dpi of the input image to ensure proper sizing. Each fingerprint is then embedded into a base block using Blender. This new fingerprint block is then printed using a Formlabs 1+ STL printer for high resolution and overall quality. After printing, the molds are cured in a homemade UV curing unit before use. 

In case of dental molds, we prepared molds using Kerr extrude (Polyinylsiloxane) impression material during the live data collection. Human live fingers were pressed into the material for approximately 4 minutes. This high quality material is used for dental applications, has very short cure time, and does not shrink.

3D molds and dental molds have the common process when making fake fingers. Once molds are prepared, we applied materials as in Table \ref{table:fingerPAdetail} on top of the molds. The quality depends on the resolution of the printer, and captured image. Dental molds generally carry better quality over 3D molds because there are no loss from data conversion. In order to increase diversity of data, the samples from a single material may differ in hardness, transparency, and color.

The last category of fake fingers is from 2D prints.  2D printing does not use molds at all. Rather, a printer applies black ink onto white paper using a drum. The source of the image is from previous fingerprint collections from traditional fingerprint scanners.  We used two types of material, paper and transparent film, from multiple printers.

We created molds from the fingerprint from right hand fingers in order to avoid taking too much time from participants.  Each presentation attack instrument was captured two times. 

In total, the dataset contains 14892 live fingerprints captures and 21700 spoof fingerprint captures from 450 subjects. 33 of the subjects are unique to live captures, 208 are unique to spoof captures, and 209 of the subjects are in both sets.  This dataset will be made available by request for comparison purposes under an IRB-approved database release agreement.

\subsubsection{Data Cleaning}

In practice, we observe a significant amount of variation in subject finger application when no capture quality feedback mechanism or guide is used during capture. Improper use of the sensor can occasionally create captures with blank frames. While a feedback mechanism and guide has been developed and implemented in a subsequent collection for this sensor, the data was not ready in time for use in this paper. To clean the data of captures with blank frames, we developed an algorithm that identifies individual blank frames using the standard deviation of the middle row of pixels. Sequences that don't contain seven sequential non-blank frames are removed from our dataset. Table \ref{table:fingerPAdetail} shows the data-set post cleaning.

\subsection{Classifier}

We evaluated the classification performance of both static and dynamic features using a fully connected Deep Neural Network (DNN) assessed using a 10-fold cross validation process. 

Two different DNNs namely \textit{DNN1}  and \textit{DNN2} are respectively evaluated on hand engineered features and features extracted using the Inception model. No pre-processing is applied to training and the features are normalized to have zero mean and unit variance. Tables \ref{DNN1} and \ref{DNN2} present the architecture of both networks. Both networks are initialized using Xavier uniform initialization \cite{glorot2010understanding} and trained using the Adam optimizer \cite{kingma2014adam} with cross entropy loss function. \textit{DNN1} employs an adaptive learning rate starting at 0.001 with patience of 10 epochs and reduce factor of 0.1 trained for 50 epochs, While \textit{DNN2} uses a fixed learning rate of 0.0005 and is trained for 60 epochs.

\begin{table}[h]
\begin{center}
\begin{tabular}{|l|c|c|c|}
\hline
Layers & Size in & Size Out & Activation\\
\hline
FC1 & 216$\times$1  & 400$\times$1 & Relu \\
\hline
FC2 & 400$\times$1  & 400$\times$1 & Relu\\
\hline
FC3 & 400$\times$1 & 2$\times$1 & Softmax \\
\hline
\end{tabular}
\end{center}
\caption{Architecture of the DNN1}
\label{DNN1}
\end{table}

\begin{table}[h]
\begin{center}
\begin{tabular}{|l|c|c|c|}
\hline
Layers & Size in & Size Out & Activation \\
\hline
FC1 & 2048$\times$1  & 500$\times$1 & Relu \\
\hline
FC2 & 500$\times$1  & 500$\times$1 & Relu \\
\hline
FC3 & 500$\times$1 & 2$\times$1 & Softmax \\
\hline
\end{tabular}
\end{center}
\caption{Architecture of the DNN2}
\label{DNN2}
\end{table}

\section{Results}
\begin{figure}[ht]
\begin{center}
  \includegraphics[width=1\linewidth]{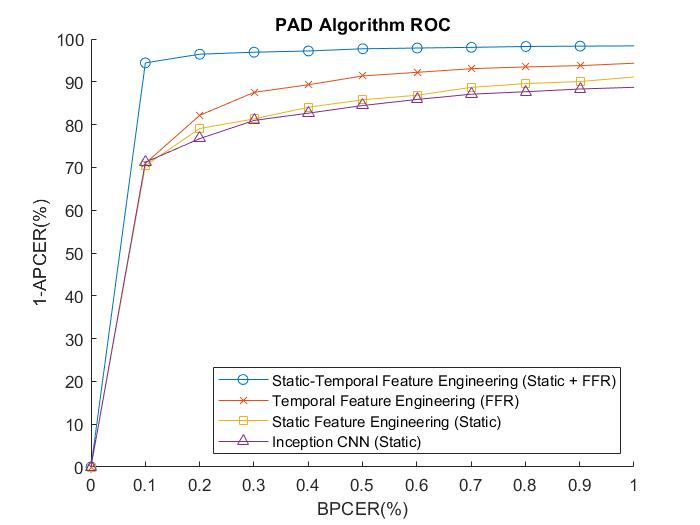}
    \caption{PAD algorithm ROC curve}
  \label{fig:results}
\end{center}
\end{figure}

In Figure \ref{fig:results} we compare the performance of static, dynamic, and fusion approaches to PAD detection using multiple ROC curves. The top curve demonstrates the performance received when deploying a classifier on top of a fusion of static and dynamic engineered features. Next down characterizes classifier performance when using solely dynamic color capture features. Further down characterizes performance from a static engineered-feature classifier. At roughly the same performance level as the former, exists our plot of classifier performance on bottleneck ImageNet-trained CNN features. The two unique methods of static feature extraction yield such similar results for this dataset. We also observe that the fusion approach yields better performance then either static or dynamic methods alone, suggesting the use of static and dynamic approaches simultaneously yields synergistic benefits. In table \ref{tab:results} we provide the mean Attack Presentation Classification Error Rate (APCER) at 0.2\% and 1.0\% Bona Fide Presentation Classification Error Rate (BPCER) operating points as well as the standard deviation (STD) at 1.0\%  BPCER.

\begin{table}[ht]
\begin{center}
\begin{tabular}{|l|c|c|c|}
\hline
PAD  & Mean  & Mean  &  STD \\
 Algorithm &  APCER     &   APCER      &     @1.0\%\\
&     @0.2\%    &   @1.0\%    &  BPCER\\
&     BPCER    &   BPCER    & \\
\hline \hline
Static-Temporal    &        &     &   \\
Engineered Features &   3.55\%    &  0.626\%    & 1.96\% \\
 (Static + FFR) &       &       & \\
\hline
Temporal Feature &        &         &      \\
Engineering (FFR) &     17.8\%     &  1.10\%       &  3.42\% \\
\hline
Static Engineered  &      &        &         \\
 Feature&  20.9\%       &   1.78\%       &  7.82\%\\
\hline
 Static Features &       &      &      \\
(Inception CNN)&  23.3\%   &    3.05\%  & 4.05\%    \\
\hline
\end{tabular}
\end{center}
\caption{PAD algorithm performance at 0.2\% BPCER and 1.0\% BPCER operating points}
\label{tab:results}
\end{table}

\section{Discussion and Conclusion}

A novel approach to fingerprint presentation attack detection using a fusion of static and color time-series fingerprint capture is detailed in this paper. When tested with a large and diverse dataset, this approach has been shown to generate classification performance greater than that of the static classifier exclusively, on a challenging dataset of diverse spoof types. Hopefully this will encourage manufacturers to bring more sensors with this technology to the marketplace and push the capabilities of spoof detection to new heights in the coming future.

The presentation attack solutions presented in this paper were designed to be generally applicable to a sensor with similar capability. Due to the lack of capability in similar sensors we could not test the algorithm's generalization to different sensors directly. 

\section{Acknowledgements}
This research is based upon work supported in part by
the Office of the Director of National Intelligence (ODNI),
Intelligence Advanced Research Projects Activity (IARPA),
via IARPA R\&D Contract No. 2017 - 17020200004. The
views and conclusions contained herein are those of the
authors and should not be interpreted as necessarily representing
the official policies, either expressed or implied,
of ODNI, IARPA, or the U.S. Government. The U.S. Government
is authorized to reproduce and distribute reprints
for governmental purposes notwithstanding any copyright
annotation therein.

{\small
\bibliographystyle{ieee}
\bibliography{egbib}

\begin{thebibliography}{10}\itemsep=-1pt

\bibitem{adair_nixon_spoof_2007}
K.~Adair~Nixon, V.~Aimale, and R.~Rowe.
\newblock Spoof {Detection} {Schemes}.
\newblock pages 403--423. Oct. 2007.

\bibitem{antonelli_fake_2006}
A.~Antonelli, R.~Cappelli, D.~Maio, and D.~Maltoni.
\newblock Fake finger detection by skin distortion analysis.
\newblock {\em IEEE Transactions on Information Forensics and Security},
  1(3):360--373, Sept. 2006.

\bibitem{cappelli_modelling_2001}
R.~Cappelli, D.~Maio, and D.~Maltoni.
\newblock Modelling {Plastic} {Distortion} in {Fingerprint} {Images}.
\newblock In S.~Singh, N.~Murshed, and W.~Kropatsch, editors, {\em Advances in
  {Pattern} {Recognition} — {ICAPR} 2001}, Lecture {Notes} in {Computer}
  {Science}, pages 371--378. Springer Berlin Heidelberg, 2001.

\bibitem{chollet2015keras}
F.~Chollet et~al.
\newblock Keras, 2015.

\bibitem{decann_novel_2009}
B.~DeCann, B.~Tan, and S.~Schuckers.
\newblock A novel region based liveness detection approach for fingerprint
  scanners.
\newblock In M.~Tistarelli and M.~S. Nixon, editors, {\em Advances in
  {Biometrics}}, Lecture Notes in Computer Science, pages 627--636. Springer
  Berlin Heidelberg, 2009.

\bibitem{deng2009imagenet}
J.~Deng, W.~Dong, R.~Socher, L.-J. Li, K.~Li, and L.~Fei-Fei.
\newblock Imagenet: A large-scale hierarchical image database.
\newblock In {\em Computer Vision and Pattern Recognition, CVPR 2009}, pages
  248--255. Ieee, 2009.

\bibitem{ghiani_textural_2015}
L.~Ghiani.
\newblock {\em Textural Features for Fingerprint Liveness Detection}.
\newblock PhD thesis, University of Cagliari, 2015.

\bibitem{glorot2010understanding}
X.~Glorot and Y.~Bengio.
\newblock Understanding the difficulty of training deep feedforward neural
  networks.
\newblock In {\em Proceedings of the thirteenth international conference on
  artificial intelligence and statistics}, pages 249--256, 2010.

\bibitem{kingma2014adam}
D.~P. Kingma and J.~Ba.
\newblock Adam: A method for stochastic optimization.
\newblock {\em arXiv preprint arXiv:1412.6980}, 2014.

\bibitem{mascaro_common_2006}
S.~Mascaro and H.~H. Asada.
\newblock The {Common} {Patterns} of {Blood} {Perfusion} in the {Fingernail}
  {Bed} {Subject} to {Fingertip} {Touch} {Force} and {Finger} {Posture}.
\newblock July 2006.

\bibitem{parthasaradhi_time-series_2005}
S.~T.~V. Parthasaradhi, R.~Derakhshani, L.~A. Hornak, and S.~A.~C. Schuckers.
\newblock Time-series detection of perspiration as a liveness test in
  fingerprint devices.
\newblock {\em IEEE Transactions on Systems, Man, and Cybernetics, Part C
  (Applications and Reviews)}, 35(3):335--343, Aug. 2005.

\bibitem{schuckers_presentations_2016}
S.~Schuckers.
\newblock Presentations and attacks, and spoofs, oh my.
\newblock {\em Image and Vision Computing}, 55:26--30, Nov. 2016.

\bibitem{szegedy2016rethinking}
C.~Szegedy, V.~Vanhoucke, S.~Ioffe, J.~Shlens, and Z.~Wojna.
\newblock Rethinking the inception architecture for computer vision.
\newblock In {\em Proceedings of the IEEE conference on computer vision and
  pattern recognition}, pages 2818--2826, 2016.

\bibitem{tan_liveness_2005}
B.~Tan and S.~Schuckers.
\newblock {\em Liveness detection using an intensity based approach in
  fingerprint scanners}.
\newblock 2005.

\bibitem{tan_liveness_2006}
B.~Tan and S.~Schuckers.
\newblock Liveness {Detection} for {Fingerprint} {Scanners} {Based} on the
  {Statistics} of {Wavelet} {Signal} {Processing}.
\newblock In {\em 2006 {Conference} on {Computer} {Vision} and {Pattern}
  {Recognition} {Workshop} ({CVPRW}'06)}, pages 26--26, June 2006.

\bibitem{yau_fake_2007}
W.-Y. Yau, H.-T. Tran, E.-K. Teoh, and J.-G. Wang.
\newblock Fake {Finger} {Detection} by {Finger} {Color} {Change} {Analysis}.
\newblock In S.-W. Lee and S.~Z. Li, editors, {\em Advances in {Biometrics}},
  Lecture {Notes} in {Computer} {Science}, pages 888--896. Springer Berlin
  Heidelberg, 2007.

\bibitem{zhang_fake_2007}
Y.~Zhang, J.~Tian, X.~Chen, X.~Yang, and P.~Shi.
\newblock Fake {Finger} {Detection} {Based} on {Thin}-{Plate} {Spline}
  {Distortion} {Model}.
\newblock In S.-W. Lee and S.~Z. Li, editors, {\em Advances in {Biometrics}},
  Lecture {Notes} in {Computer} {Science}, pages 742--749. Springer Berlin
  Heidelberg, 2007.

\end{thebibliography}
}

\end{document}